\PassOptionsToPackage{table}{xcolor}
\documentclass[sigconf]{acmart}

\newsavebox\CBox

\definecolor[named]{BoxColor}{RGB}{224,234,248}
\definecolor[named]{lightred}{RGB}{255,222,219}
\definecolor[named]{lightgreen}{RGB}{234,255,233}

\copyrightyear{2024}
\acmYear{2024}
\setcopyright{rightsretained}
\acmConference[RecSys '24]{18th ACM Conference on Recommender Systems}{October 14--18, 2024}{Bari, Italy}
\acmBooktitle{18th ACM Conference on Recommender Systems (RecSys '24), October 14--18, 2024, Bari, Italy}\acmDOI{10.1145/3640457.3691719}
\acmISBN{979-8-4007-0505-2/24/10}

\title{Enhancing Sequential Music Recommendation with Personalized Popularity Awareness}

\author{Davide Abbattista}
\affiliation{
  \institution{Politecnico di Bari, Italy}
  \country{}
}
\email{d.abbattista@studenti.poliba.it}

\author{Vito Walter Anelli}
\affiliation{
  \institution{Politecnico di Bari, Italy}
  \country{}
}
\email{vitowalter.anelli@poliba.it}

\author{Tommaso Di Noia}
\affiliation{
  \institution{Politecnico di Bari, Italy}
  \country{}
}
\email{tommaso.dinoia@poliba.it}

\author{Craig Macdonald}
\affiliation{
  \institution{University of Glasgow, UK}
  \country{}
}
\email{craig.macdonald@glasgow.ac.uk}

\author{Aleksandr Petrov}
\affiliation{
  \institution{University of Glasgow, UK}
  \country{}
}
\email{a.petrov.1@research.gla.ac.uk}

\renewcommand{\shortauthors}{Abbattista et al.}



\usepackage{multirow}
\usepackage{balance}
\usepackage[noadjust]{marginnote}

\newcommand{\crc}[1]{\textcolor{red}{#1}}

\begin{document}

\begin{abstract}
In the realm of music recommendation, sequential recommender systems have shown promise in capturing the dynamic nature of music consumption. Nevertheless, traditional Transformer-based models, such as SASRec and BERT4Rec, while effective, encounter challenges due to the unique characteristics of music listening habits. In fact, existing models struggle to create a coherent listening experience due to rapidly evolving preferences. Moreover, music consumption is characterized by a prevalence of repeated listening, i.e. users frequently return to their favourite tracks, an important signal that could be framed as individual or personalized popularity. This paper addresses these challenges by introducing a novel approach that incorporates personalized popularity information into sequential recommendation. By combining user-item popularity scores with model-generated scores, our method effectively balances the exploration of new music with the satisfaction of user preferences.
Experimental results demonstrate that a Personalized Most Popular recommender, a method solely based on user-specific popularity, outperforms existing state-of-the-art models.
Furthermore, augmenting Transformer-based models with personalized popularity awareness yields superior performance, showing improvements ranging from 25.2\% to 69.8\%. 
The code for this paper is available at \url{https://github.com/sisinflab/personalized-popularity-awareness}.
\end{abstract}

\maketitle

\section{Introduction}
Sequential recommendation allows the prediction of potential items of interest to users based on their past interactions. It is a commonly deployed form of recommendation that takes into account the order of interactions when making predictions. Indeed, in many recommendation domains such as e-commerce, music, and POI, the order of recent interactions can provide more fine-grained evidence of users' current interests and context than the traditional set-based collaborative filtering mechanism \cite{DBLP:journals/csur/QuadranaCJ18}.

In our work, we focus on music recommendation. The music domain exhibits unique characteristics such as vast catalog sizes, repeated consumption patterns, and sequential and passive consumption behaviors, each posing distinct challenges for recommender systems \cite{DBLP:reference/sp/SchedlKMB22}.
Music listening habits are inherently repetitive. Therefore, a recommender system must identify and prioritize users' favorite tracks to maintain their satisfaction and engagement. Ignoring previously enjoyed tracks can lead to a disconnect between the user’s preferences and the recommendations, reducing the system’s perceived relevance. 

Transformer-based models \cite{DBLP:conf/nips/VaswaniSPUJGKP17} are widely adopted for sequence recommendation due to their popularity in other sequential tasks such as language modeling. Indeed, sequence recommendation is similar to the language modeling task of sentence completion if items are treated like language tokens. Models such as SASRec \cite{DBLP:conf/icdm/KangM18} (and its recent adaption, gSASRec \cite{DBLP:conf/recsys/PetrovM23}) and BERT4Rec \cite{DBLP:conf/cikm/SunLWPLOJ19} have shown to be effective for sequence recommendation and have been used as baselines for music recommendation in previous works, such as \cite{DBLP:conf/sigir/TranSSH23, DBLP:conf/cikm/ZhouWZZWZWW20, DBLP:conf/cikm/FanLZX0Y21, DBLP:conf/ijcai/TongWLXN21, DBLP:conf/aaai/Shin0WP24, DBLP:conf/dis/AmjadiTT21, DBLP:journals/ijcrowdsci/ZhangC0LW21}. However, as we show later in this paper, music recommendation presents significant complexity for such Transformer-based models due to its inherent characteristics, especially the repeated consumption patterns.

Our take on the problem is:
Giving knowledge of the personalized popularity of all track items for each user to sequential models can help them to learn more refined user-specific recommendations due to the repeated consumption patterns in the music domain. Although popularity is often used as a baseline~\cite{DBLP:conf/ecir/AnelliNSRT19,DBLP:conf/aaai/RenCLR0R19,DBLP:conf/sigir/AriannezhadJLFS22,DBLP:journals/tois/LiJAR23}, this is the first approach directly injecting a personalized popularity signal into sequential models.

The contributions of this work are as follows:
(1) A novel approach to integrate personalized popularity awareness into sequential recommender systems; (2) an experimental evaluation of the proposed approach.
Our results show that (i) \textit{Personalized Most Popular} recommender, which ranks track items according to user-specific popularity, has the best performance compared to state-of-the-art sequential recommenders, (ii) combining personalized popularity scores within Transformer-based models improves the quality of music recommendation, beating \textit{Personalized Most Popular} recommender at high ranking cutoffs.
\section{Personalized Popularity Awareness}\label{sec:ppa}
This section introduces the concept of \textit{personalized popularity awareness} and its application in music recommendation.
Personalized popularity refers to the tendency of users to repeatedly listen to tracks they have enjoyed in the past. This behaviour, known as repeated consumption, is a distinctive characteristic of music consumption. Indeed, \textit{Personalized Most Popular} is an often-employed baseline, which recommends items based on user-specific popularity, and has been used as a baseline in previous works on time-aware kNN~\cite{DBLP:conf/ecir/AnelliNSRT19}, session-based~\cite{DBLP:conf/aaai/RenCLR0R19, DBLP:journals/corr/HidasiKBT15} and next basket~\cite{DBLP:conf/kdd/HuH19, DBLP:conf/sigir/AriannezhadJLFS22, DBLP:conf/um/FaggioliPA20, DBLP:journals/tois/LiJAR23} recommendation, outperforming, in domains with high repetitiveness, state-of-the-art models or being in the same range as them. 
In addition, previous works on sequential~\cite{DBLP:conf/dasfaa/MaZLSXZ20, DBLP:conf/cikm/QuanLZW23} and session-aware~\cite{DBLP:conf/iccpr/YangZL20} recommendation have introduced novel architectures to model the repeated consumption pattern.
To the best of our knowledge, this is the first attempt at directly injecting a personalized popularity signal into sequential models.

\looseness -1 We argue that traditional recommendation models, such as Trans\-former-based models like SASRec and BERT4Rec, often struggle to effectively capture this behavior. To address this, we introduce a novel approach that integrates personalized popularity information directly into sequential recommendation models. This is achieved by combining user-item popularity scores with model-generated scores assigned to items (music tracks) by different recommendation models. In essence, instead of learning the entire probability distribution of item preferences, our key insight is that the training process for models like BERT4Rec, SASRec and gSASRec can be adapted to focus on {\em deviations} from the popularity distribution.
This intuition is similar to gradient boosting, where a model is tasked with learning the delta from a previous model~\cite[Ch.10]{hastie01statisticallearning}. In our case, the models are tasked with learning the delta from personalized popularity, i.e. recommendation going beyond the repeated consumed tracks. 

\vspace{0.5em}
\noindent \textbf{Sequential Models Item Probability.}
Considering the user input sequence of item ids \(S=[s_1,\ldots,s_i,\ldots,s_L]\), where \(L\) is the maximum input sequence length, let \(L_i=[x_{i_1},\ldots,x_{i_j},\ldots,x_{i_N}]\) be the scores vector corresponding to \(s_i\) output by a sequential recommendation model, such as BERT4Rec or SASRec, where \(x_{i_j}\) is the score of the item \(j\) related to the position \(i\) and \(N\) is the number of items in the catalogue.
The loss function used for the training of BERT4Rec model is Cross Entropy, where 
\begin{equation}\label{eq:softmax}
    p_M(j_i)=softmax(x_{j_i})=\frac{e^{x_{j_i}}}{\sum_{z=1}^{N} e^{x_{j_z}}}
\end{equation}
is the probability of item \(j\) to be ranked in position \(i\). Similarly, the loss used for the training of SASRec and gSASRec models is the Binary Cross Entropy (BCE) and the generalized Binary Cross Entropy loss (gBCE), respectively, where
\begin{equation}\label{eq:sigmoid}
    p_M(j_i)=sigmoid(x_{j_i})=\frac{1}{1+e^{-x_{j_i}}}
\end{equation}
is the probability of item \(j\) to at position \(i\) (in case of gBCE, the $sigmoid(\cdot)$ is then raised to the power of $\beta$ before applying the loss). 

\vspace{0.5em}
\noindent \textbf{Personalized Popularity Item Probability.} 
Let us denote with \(C=[c_{1},\ldots,c_{i},\ldots,c_{N}]\) the {\em counts vector} obtained from the input sequence \(S\), where \(c_i\) is the count of item \(i\) and \(N\) is the number of items in the catalogue. The probability of a previous item \(j\) being selected by the user, i.e. its {\em personalized popularity} is\footnote{We divide by $max(C)$ to avoid numerical instabilities.}:
\begin{equation}\label{eq:pp}
    p_P(j)=\frac{c_j}{\sum_{z=1}^{N} c_z}=\frac{\frac{c_j}{max(C)}}{\sum_{z=1}^{N} \frac{c_z}{max(C)}}.
\end{equation}

We further smooth $ p_P(j)$ using an \(\epsilon >0\) value, which can be tuned in order to control the contribution of the personalized popularity on the score of each item, once the model's score and the personalized popularity are combined. In our experiments \(\epsilon\) is set to 0.01. By smoothing, the personalized popularity probability of item \(j\) becomes: 
\begin{equation}\label{eq:ppeps}
    \hat{p}_P(j)=\frac{\frac{c_j+\epsilon}{max(C+\epsilon)}}{\sum_{z=1}^{N} \frac{c_z+\epsilon}{max(C+\epsilon)}}.
\end{equation}
\looseness -1 The lower is \(\epsilon\), the higher the contribution of personalized popularity on the overall items' score. As \(\epsilon\rightarrow0\)\ the probability of item $j$ corresponds to the unsmoothed personalized popularity probability $p_P(j)$ in Eq.~\eqref{eq:pp}. As \(\epsilon\rightarrow+\infty\), personalized popularity item probabilities have the same value, for all items in the catalogue. As a consequence, with \(\epsilon\rightarrow+\infty\) the personalized popularity will have less influence on the item ranking. Combination of $\hat{p}_P(j)$ with model outputs depends on the activation function used by the model's loss function (sigmoid or softmax), as we explain next.

\vspace{0.5em}
\noindent \textbf{Using Personalized Popularity in a Softmax.}
In BERT4Rec, the probability of an item is obtained using the softmax function (Eq.~\eqref{eq:softmax}). Therefore, to combine $\hat{p}_P(j)$ into the softmax, we need to transform the $\hat{p}_P(j)$ into a compatible score, which we denote $y_j$, that can be combined with the model's score. By imposing an equivalence between Eq.~\eqref{eq:softmax} and Eq.~\eqref{eq:ppeps}, the personalized popularity score $y_j$ of an item $j$ is the exponent value in Eq.~\eqref{eq:softmax}. Therefore, to inject  the personalized popularity probability into the model's scoring, we can combine Eq.~\eqref{eq:softmax} and Eq.~\eqref{eq:ppeps} as follows:
\label{subsec:pers_pop_logits_softmax}
\begin{equation}\label{eq:softmax_pp}
    \hat{p}_P(j)=\frac{\frac{c_j+\epsilon}{max(C+\epsilon)}}{\sum_{z=1}^{N} \frac{c_z+\epsilon}{max(C+\epsilon)}}=\frac{e^{y_j}}{\sum_{z=1}^{N} e^{y_z}}.
\end{equation}
Next, the personalized popularity probability can be formulated as a softmax following the equivalence
\begin{equation}\label{eq:exp_pp}
    \frac{\frac{c_j+\epsilon}{max(C+\epsilon)}}{\sum_{z=1}^{N} \frac{c_z+\epsilon}{max(C+\epsilon)}}=\frac{e^{\ln{\frac{c_j+\epsilon}{max(C+\epsilon)}}}}{e^{\ln{\sum_{z=1}^{N} \frac{c_z+\epsilon}{max(C+\epsilon)}}}}.
\end{equation}
Combining Eq.~\eqref{eq:softmax_pp} with Eq.~\eqref{eq:exp_pp}, we obtain
\[
    \frac{e^{\ln{\frac{c_j+\epsilon}{max(C+\epsilon)}}}}{e^{\ln{\sum_{z=1}^{N} \frac{c_z+\epsilon}{max(C+\epsilon)}}}}=\frac{e^{y_j}}{\sum_{z=1}^{N} e^{y_z}}.
\]
Therefore we can formulate the personalized popularity score of the item \(j\) for BERT4Rec as
\begin{equation}
    y_j=\ln{\frac{c_j+\epsilon}{max(C+\epsilon)}}.
\end{equation}

To integrate personalized popularity score into the BERT4Rec model, starting from the user input sequence \(S\), we compute the personalized popularity scores vector \(P=[y_1,\ldots,y_j,\) \(\ldots,y_N]\), and sum it to each scores vector \(L_i\) obtained from BERT4Rec, where \(i\) is the position in the sequence.

\vspace{0.5em}
\noindent \textbf{Using Personalized Popularity in a Sigmoid.}
\label{subsec:pers_pop_logits_sigmoid}
In SASRec (and gSASRec), the probability of an item $j$ is obtained using the sigmoid function. Consequently, the personalized popularity score $y_j$ of an item $j$ is the additive inverse of the exponent value in the sigmoid formula (Eq.~\eqref{eq:sigmoid}). Combining Eq.~\eqref{eq:sigmoid} with Eq.~\eqref{eq:ppeps} gives:
\begin{equation}\label{eq:sigmoid_pp}
    \hat{p}_P(j)=\frac{\frac{c_j+\epsilon}{max(C+\epsilon)}}{\sum_{z=1}^{N} \frac{c_z+\epsilon}{max(C+\epsilon)}}=\frac{1}{1+e^{-y_j}}.
\end{equation}
This implies that $y_j$ can be isolated in Eq.~\eqref{eq:sigmoid_pp} as follows: 
\[
\begin{split}
    &\hat{p}_P(j)\cdot (1+e^{-y_j})=1\Rightarrow \hat{p}_P(j)+\hat{p}_P(j)\cdot e^{-y_j}=1\\&\Rightarrow e^{-y_j}=\frac{1-\hat{p}_P(j)}{\hat{p}_P(j)}\Rightarrow y_j=-\ln{\frac{1-\hat{p}_P(j)}{\hat{p}_P(j)}}.
\end{split}
\]
To integrate personalized popularity scores into the SASRec and gSASRec models, starting from the user input sequence \(S\), we compute a personalized popularity scores matrix \(P\) and sum it to the scores matrix \(L\) obtained from the model. We define a vector of personalized popularity scores as \(P_i=[y_{i_1},\ldots,y_{i_j},\ldots,y_{i_N}]\), where \(i\) is the position in the sequence, after removing all the items after position \(i\) from the input sequence \(S\). This ensures that the popularity scores at each position are computed without experiencing temporal leakage from future data, and therefore maintain the causal structure of the sequence.
\section{Experimental Setup}
\label{sec:exp_setup}
\begin{table}
\setlength{\tabcolsep}{2pt}
\caption{Key statistics for the two music datasets, before and after applying a popularity-based sampling method to select 30,000 items (\textit{N}). \textit{Avg len} and \textit{Med len} denote the average and the median number of interactions per user, respectively.}\vspace{-1em}
\footnotesize
\centering
\begin{tabular}{lcccccc}
    \toprule
    Dataset& Sampling (\(N=30000\))& Users& Items& Interactions& Avg. len& Med. len\\ 
    \midrule
    \multirow{2}{5em} {Yandex}& \cellcolor{gray!20}None&\cellcolor{gray!20}20862&\cellcolor{gray!20}300000&\cellcolor{gray!20}46919149&\cellcolor{gray!20}2249&\cellcolor{gray!20}2289\\
    &Popularity-based &20862&30000&28408152&1362&1275\\
    \midrule
    \multirow{2}{5em}{Last.fm-1K}&\cellcolor{gray!20}  None&\cellcolor{gray!20}992&\cellcolor{gray!20}961416&\cellcolor{gray!20}16982280&\cellcolor{gray!20}17119&\cellcolor{gray!20}10249\\
    &Popularity-based &990&30000&4990042&5040&2605\\
    \bottomrule
\end{tabular}\vspace{-1em}
\label{tab:dataset}
\end{table}
\vspace{0.5em}
\noindent \textbf{Research Questions.} This section examines the effectiveness of incorporating personalized popularity awareness into sequential music recommender systems through a series of experiments designed to answer the following research questions:
\begin{enumerate}
    \item[\textbf{RQ1}] How effectively do state-of-the-art sequential recommender models behave on music datasets?  
    \item[\textbf{RQ2}] To what extent does explicitly incorporating personalized popularity information enhance the performance of sequential recommendation models in the context of music?
\end{enumerate}

\looseness -1 \noindent \textbf{Datasets.}
Our experiments are performed on the Yandex music event dataset\footnote{\url{https://www.kaggle.com/competitions/yandex-music-event-2019-02-16}} and on the Last.fm-1K dataset\footnote{\url{http://ocelma.net/MusicRecommendationDataset/lastfm-1K.html}} \cite{Celma:Springer2010}. 
In the Yandex dataset from 2019, there are four types of time-stamped interaction events on music tracks: \textit{like}, \textit{dislike}, \textit{play} and \textit{skip}. 
Last.fm-1K is a smaller dataset, published in 2010, containing time-stamped play events for  nearly 1,000 users. 
Since training models with the whole datasets would require a large amount of GPU memory, we use a popularity-based item sampling method, which selects $N$ items based on a probability distribution derived from global items counts in the users interactions, thus ensuring that more popular items have a higher chance of being selected while still allowing for the inclusion of less popular items. Other sampling methods, such as uniformly sampling interactions, would have not given the same guarantees of not injecting an additional popularity bias. Table 1 reports the overall statistics of the datasets before and after item sampling.

\vspace{0.5em}
\noindent \textbf{Models.}
The recommender models considered in the experiments comprise two variants of the Most Popular recommender and some of the most known sequential Transformer-based recommenders both in their original formulation, and integrating the personalized popularity scores (PPS) as per Section~2.
The \textit{Most Popular} recommender suggests items based on overall popularity across all users. Although it is a strong baseline~\cite{DBLP:conf/recsys/CremonesiKT10}, in experiments we see \textit{Most Popular} exhibits sub-optimal performance compared to other models. We further include \textit{Personalized Most Popular}, following many existing works~\cite{DBLP:conf/ecir/AnelliNSRT19,DBLP:conf/aaai/RenCLR0R19,DBLP:journals/corr/HidasiKBT15,DBLP:conf/kdd/HuH19, DBLP:conf/sigir/AriannezhadJLFS22, DBLP:conf/um/FaggioliPA20, DBLP:journals/tois/LiJAR23}. This baseline leverages the tendency of users to prefer and repeatedly listen to previously enjoyed music, demonstrating robust performance and highlighting the repeated consumption characteristic of music recommendation. 

To evaluate the effectiveness of the \textit{Personalized Most Popular} baseline and personalized popularity score, we also compare them with state-of-the-art models for sequential recommendation: BERT4Rec \cite{DBLP:conf/cikm/SunLWPLOJ19}, SASRec \cite{DBLP:conf/icdm/KangM18}, and its recent variant gSASRec \cite{DBLP:conf/recsys/PetrovM23}. BERT4Rec is a model for sequential recommendations that leverages bidirectional self-attention mechanisms. SASRec is a sequential recommendation model that uses unidirectional self-attention. gSASRec (generalized SASRec) is a version of SASRec model with an increased number of negatives, trained with gBCE loss. 
Finally, to show the effectiveness of personalized popularity score, we integrate it into these models, assessing whether they generate more accurate and relevant music recommendations.
\begin{table*}
\setlength{\tabcolsep}{3pt}
\footnotesize
\centering
\begin{tabular}{lllllllll}
    \toprule
    \multirow{3}{*}{Model}&\multicolumn{4}{c}{Yandex}&\multicolumn{4}{c}{Last.fm-1K}\\
    \cmidrule(lr){2-5}\cmidrule(lr){6-9}
    &NDCG@5&NDCG@10&NDCG@40&NDCG@100&NDCG@5&NDCG@10&NDCG@40&NDCG@100\\
    \midrule
    \rowcolor[gray]{.9} Most Popular&0.0447&0.0416&0.0386&0.0449&0.0946&0.0889&0.0831&0.0817\\
    
    Personalized Most Popular&\textbf{0.1866}*&\textbf{0.1826}*&0.1721&0.1947&\textbf{0.5366}*&\textbf{0.5056}*&\textbf{0.4138}&\underline{0.3554}\\
    \midrule
    \rowcolor[gray]{.9} BERT4Rec&0.1311&0.1263&0.1248&0.1466&0.3504&0.3164&0.2437&0.2108\\
    BERT4Rec w/ PPS&\underline{0.1780}\dag\hspace{0.1em}(+35.8\%)&\underline{0.1745}\dag\hspace{0.1em}(+38.2\%)&\textbf{0.1765}*\dag\hspace{0.1em}(+41.4\%)&\textbf{0.2024}*\dag\hspace{0.1em}(+38.1\%)&\underline{0.5096}\dag\hspace{0.1em}(+45.4\%)&\underline{0.4812}\dag\hspace{0.1em}(+52.1\%)&\underline{0.4100}\dag\hspace{0.1em}(+68.2\%)&\textbf{0.3579}\dag\hspace{0.1em}(+69.8\%)\\
    \midrule
    \rowcolor[gray]{.9} SASRec&0.1025&0.1030&0.1086&0.1332&0.3381&0.3160&0.2625&0.2317\\
    SASRec w/ PPS&0.1658\dag\hspace{0.1em}(+61.8\%)&0.1647\dag\hspace{0.1em}(+59.9\%)&0.1718\dag\hspace{0.1em}(+58.2\%)&0.2012\dag\hspace{0.1em}(+51.1\%)&0.4500\dag\hspace{0.1em}(+33.1\%)&0.4258\dag\hspace{0.1em}(+34.8\%)&0.3785\dag\hspace{0.1em}(+44.2\%)&0.3410\dag\hspace{0.1em}(+47.2\%)\\
    \midrule
    \rowcolor[gray]{.9} gSASRec&0.1418&0.1364&0.1353&0.1592&0.3475&0.3222&0.2559&0.2218\\
    gSASRec w/ PPS&0.1776\dag\hspace{0.1em}(+25.2\%)&0.1737\dag\hspace{0.1em}(+27.3\%)&\underline{0.1747}\dag\hspace{0.1em}(+29.1\%)&\underline{0.2013}\dag\hspace{0.1em}(+26.4\%)&0.4990\dag\hspace{0.1em}(+43.6\%)&0.4710\dag\hspace{0.1em}(+46.2\%)&0.4006\dag\hspace{0.1em}(+56.6\%)&0.3534\dag\hspace{0.1em}(+59.3\%)\\
    \bottomrule
\end{tabular}
\caption{Experimental results. 
The best results are highlighted in bold; second best are underlined. * and \dag \hspace{0.1em} denote statistically significant differences (\textit{p-value} < 0.05, Bonferroni multi-test correction) between the best model and the second best one, and between the model w/ PPS and the one w/o PPS., respectively.  
Improvements over models w/o PPS are shown in parentheses.}\vspace{-3em}
\label{tab:results}
\end{table*}

\vspace{0.5em}
\noindent \textbf{Implementation Details.}
We use PyTorch\footnote{\url{https://pytorch.org/docs/stable/index.html}} \cite{DBLP:conf/nips/PaszkeGMLBCKLGA19} library to implement all the models, except for \textit{Most Popular} and \textit{Personalized Most Popular}. Moreover, we rely on Hugging Face's Transformers\footnote{\url{https://huggingface.co/transformers}} \cite{DBLP:conf/emnlp/WolfDSCDMCRLFDS20} library to implement BERT4Rec with and without PPS. We use the aprec\footnote{\url{https://github.com/asash/BERT4Rec\_repro}} framework from \cite{DBLP:conf/recsys/PetrovM22a}.
We selected the hyperparameters, i.e. the choice of optimizer (Adam, AdamW), the learning rate (0.1, 0.001, 0.0001), the weight decay (0, 0.5), the early stopping patience (200, 500) and the $\epsilon$ smoothing parameter (0.01, 0.05, 0.1, 0.2), through an extensive search ($\approx338$ GPU hours). For further research we made the chosen hyperparameters and code publicly available\footnote{\url{https://github.com/sisinflab/personalized-popularity-awareness}}.
Experiments are conducted using 16-cores of an AMD Ryzen Threadripper PRO 3955WX CPU with 128GB of RAM and an Nvidia RTX 3090 GPU with 24GB of VRAM. Given these capabilities, we limit the length of the input sequences to 150.

\vspace{0.5em}
\noindent \textbf{Evaluation Details.}
\looseness -1 Following sequential recommendation best practices \cite{DBLP:conf/icmcs/ChouYL15, DBLP:conf/recsys/HidasiC23a, DBLP:reference/sp/SchedlKMB22}, we evaluate models using a Global Temporal Split strategy and Normalized Discounted Cumulative Gain (NDCG) \cite{DBLP:journals/tois/JarvelinK02}.
The Global Temporal Split strategy splits the interactions in the dataset given a test fraction and a validation fraction, both set to \(0.1\).
Specifically, all interactions in the dataset are sorted by their timestamp and the global test border timestamp is obtained taking the one of the interactions in the position that leaves 10\% of all interactions for testing. 
For each user we hold out the interactions after the global test border timestamp as the test set.
We also construct a validation set with the same strategy using a group of 2048 users with Yandex music event dataset and 128 users with Last.fm-1K dataset (approximately the \(10\%\) of the users). 
Following \citet{DBLP:conf/iui/YakuraNG18, DBLP:journals/umuai/YakuraNG22}, in the Yandex dataset, we assign labels to items for the purposes of calculating NDCG, based on the user-track interaction type: like=2, play=1\crc, skip=-1, dislike=-2. To handle the repeated consumption patterns within the dataset ground truth, the label is assigned based on the first interaction and updated only if the absolute value of the new one is greater or equal than the absolute value of the previous one:  (i) if a user likes or dislikes a track, the corresponding label is updated, regardless its past interactions with that track, (ii) if a user plays or skips a track, the corresponding label is updated only if that user has not liked nor disliked that track in the past.
Negative interactions (dislike, skip) are then assigned a label of 0, to avoid instability of NDCG in the presence of negatively scored labels \cite{DBLP:conf/cikm/GienappFHP20}.
\section{Results}
\label{sec:results}
Our experimental results are presented in Table~\ref{tab:results}, which compares the performance of various recommendation models on the Yandex and Last.fm-1K datasets. These models include variations of Most Popular recommender, Transformer-based models, and their counterparts enhanced with personalized popularity scores (PPS).
To measure the significance of performance differences, we use the paired t-test with Bonferroni multiple testing correction (\textit{p-value} < 0.05), following recommended practices in IR \cite{DBLP:journals/sigir/Fuhr20}.
In the table, we use * to denote statistically significant differences between the best and second-best models, and $\dag$ to denote  statistically significant differences between models with and without integrated personalized popularity scores.
The table presents the results using the NDCG at various cutoffs (@5, @10, @40, and @100). Additionally, the relative improvements achieved by incorporating personalized popularity scores are shown in parentheses.

\vspace{0.5em}
\looseness -1 \noindent \textbf{RQ1.}
With respect to the performance of state-of-the-art transformer-based sequential recommenders, we unexpectedly find that BERT4Rec, SASRec, and gSASRec fail to outperform \textit{Personalized Most Popular} in the context of music recommendation. Although the strength of this baseline is known in the literature \cite{DBLP:conf/aaai/RenCLR0R19, DBLP:journals/corr/HidasiKBT15, DBLP:conf/kdd/HuH19, DBLP:conf/sigir/AriannezhadJLFS22, DBLP:conf/um/FaggioliPA20, DBLP:journals/tois/LiJAR23, DBLP:conf/sigir/Hu0GZ20}, its dominance in the music domain has not been previously highlighted. Our results underscore the importance of user-specific popularity in music recommendation and indicate that tracks frequently enjoyed by users have a high probability of being replayed. This highlights the strong repetitive consumption patterns among listeners.
In addition, our results reveal a significant challenge faced by state-of-the-art sequential recommendation models, such as BERT4Rec, SASRec, and gSASRec. These models struggle to effectively capture and learn the repeated consumption patterns that are essential for providing accurate and relevant music recommendations. This difficulty suggests a need for further refinement and adaptation of these models to better address the characteristics of music listening behavior. 

\vspace{0.1cm}
\noindent \setlength{\fboxsep}{10pt}
\setlength{\fboxrule}{2pt}
\hspace*{-1.1em}%
\fcolorbox{white}{BoxColor}{%
    \parbox{0.97\columnwidth}{%
        Contrary to expectations, the more complex Transformer-based models did not outperform the Personalized Most Popular recommender, revealing that users' strong preference for replaying previously enjoyed tracks is a more significant factor in music recommendation than complex algorithmic predictions.
        }
    }
    \vspace{0.1cm}


\looseness -1 \noindent \textbf{RQ2.}
By incorporating personalized popularity scores (PPS) into Transformer-based models, we achieve performance levels comparable to the simplistic \textit{Personalized Most Popular} baseline at lower ranking cutoffs, such as NDCG@10, and surpass it at higher ranking cutoffs, such as NDCG@100. This integration allows the models to leverage the inherent user-specific popularity bias effectively.
Our results indicate that introducing an inclination towards frequently played tracks significantly enhances model performance, in a range from 25.2\% to 69.8\%. This highlights the importance of accounting for users' tendencies to repeatedly enjoy certain tracks, suggesting that Transformer-based models can benefit greatly from integrating personalized popularity information. By doing so, these models become more adept at reflecting the natural listening habits of users, leading to more effective and satisfying music recommendations.

    \vspace{0.1cm}
\noindent \setlength{\fboxsep}{10pt}
\setlength{\fboxrule}{2pt}
\hspace*{-1.1em}%
\fcolorbox{white}{BoxColor}{%
    \parbox{0.97\columnwidth}{%
        By explicitly integrating personalized popularity information into Transformer-based models, their performance in music recommendation was significantly improved, achieving comparable or even surpassing the Personalized Most Popular recommender, demonstrating the importance of directly incorporating users' preference for replaying previously enjoyed tracks.
        }
    }
    \vspace{0.1cm}

\section{Conclusions}
\label{sec:conclusions}
This work demonstrates that music recommendation benefits considerably from incorporating personalized popularity awareness. Experimental results on the Yandex Music Event and Last.fm-1K datasets show that recommending based on user-specific listening history, as exemplified by the Personalized Most Popular recommender, outperforms several state-of-the-art sequential models, including BERT4Rec, SASRec, and gSASRec. This highlights the significant influence of repeated listening patterns in music consumption, which the more complex models struggle to capture effectively.
Furthermore, integrating personalized popularity scores into Transformer-based models leads to performance comparable to the Personalized Most Popular recommender at lower ranking cutoffs and even surpasses it at higher cutoffs. This suggests that directly incorporating a bias towards previously enjoyed tracks enhances recommendation accuracy.
Future work can explore several promising avenues, including investigating techniques beyond scores integration, exploring the impact of popularity awareness in other recommendation domains, and examining the long-term effects of popularity-aware recommendation, specifically the potential to create filter bubbles or limit user exposure to novel content. 

\begin{acks}
Anelli and Di Noia acknowledge partial support of the following projects: OVS: Fashion Retail Reloaded, Lutech Digitale 4.0, Secure Safe Apulia, Patti Territoriali WP1, BIO-D, Reach-XY. 
\end{acks}
\balance
\bibliographystyle{ACM-Reference-Format}
\bibliography{references}

\end{document}